\documentclass[aps,prl,twocolumn,footinbib]{revtex4-1}
\usepackage{amsmath, mathrsfs, amssymb,amsfonts,amsthm,graphicx, epsf, dcolumn, yfonts}
\usepackage[hyperfootnotes=true]{hyperref}
\usepackage{subfigure}
\usepackage{color}
\usepackage{slashed}
\usepackage{setspace}
\usepackage{cancel}
\usepackage{wasysym}
\pdfoutput=1
 \newcommand{\be}{\begin{equation}}
 \newcommand{\ee}{\end{equation}}
 \newcommand{\bea}{\begin{eqnarray}}
 \newcommand{\eea}{\end{eqnarray}}

\def\p{\partial}

\renewcommand*{\thefootnote}{\fnsymbol{footnote}}

\begin{document}

\title{Chaotic strings in AdS/CFT}
\author{Jan de Boer,$^\dag$ Eva Llabr\'es,$^\dag$ Juan F. Pedraza,$^\dag$ and David Vegh$^\star$}
%\email{J.deBoer@uva.nl,  e.m.llabres@uva.nl,  jpedraza@uva.nl, d.vegh@uu.nl}
\affiliation{\vspace{1mm}$^\dag$Institute for Theoretical Physics, University of Amsterdam,
 1090 GL Amsterdam, Netherlands\\
$^\star$Institute for Theoretical Physics, Utrecht University, 3584 CC Utrecht, Netherlands}

\begin{abstract}\vspace{-2mm}
Holographic theories with classical gravity duals are maximally chaotic; i.e., they saturate
the universal bound on the rate of growth of chaos \cite{Maldacena:2015waa}. It is interesting to ask
whether this property is true only for leading large $N$ correlators or if it can show up elsewhere.
In this Letter we consider the simplest setup to tackle this question: a Brownian particle coupled to a thermal ensemble. We find that
the four-point out-of-time-order correlator that diagnoses chaos initially grows at an exponential rate that saturates the chaos bound, i.e., with a Lyapunov exponent $\lambda_L=2\pi/\beta$. However, the scrambling time is parametrically smaller than for plasma excitations, $t_*\sim\beta \log \sqrt{\lambda}$ instead of $t_*\sim\beta \log N^2$. Our result shows that, at least in certain cases, maximal chaos can be attained in the probe sector without the explicit need of gravitational degrees of freedom.
\end{abstract}

\renewcommand*{\thefootnote}{\arabic{footnote}}
\setcounter{footnote}{0}

\maketitle

\noindent \textbf{1. Introduction.}
In recent years, the study of quantum chaos in AdS/CFT
has become a topic of great interest, leading to
new insights in quantum gravity and conformal
field theories. This program was initiated in
\cite{Shenker:2013pqa} which presented the first holographic realization
of the butterfly effect. More recently, the same approach has been
generalized to various other gravitational setups \cite{Shenker:2014cwa,ButterflyOther}.

In quantum mechanical systems, one way to analyze chaos is through the commutator $[W(t),V(0)]$ between a pair of Hermitian operators. This commutator represents the sensitivity of $W(t)$ to perturbations created at an initial time by $V(0)$. The strength of this effect is measured by the quantity
\be\label{coft}
C(t)=-\langle[W(t),V(0)]^2\rangle,
\ee
where the bracket denotes a thermal expectation value at temperature $T=\beta^{-1}$.  The time at which $C(t)$ becomes significant is called the scrambling time $t_*$. The quantity $C(t)$ contains time-ordered and out-of-time-ordered correlators. Time-ordered correlators are not sensitive to chaos: they decay as $\langle V(0)V(0) W(t) W(t) \rangle \sim \langle V V \rangle \langle W W \rangle + {\cal O}(e^{-t/t_d})$, where $t_d\sim \beta$ is the dissipation time.  The chaotic behavior of (\ref{coft}) can be probed by the out-of-time-order correlator (OTOC)
\be\label{otoc}
f(t)=\frac{\langle V W(t) V W(t)\rangle}{\langle V V\rangle\langle W W\rangle},
\ee
which becomes small at late times if the system is chaotic.
For instance, in holographic theories with Einstein gravity duals, one finds that, for $t_d<t<t_*$ \cite{Shenker:2013pqa,Shenker:2014cwa,ButterflyOther},
\be\label{fgravity}
f(t)=1-\frac{f_0}{N^2}e^{\lambda_L t}+\mathcal{O}(N^{-4}),
\ee
where $f_0$ is a positive order one constant that depends on the specific operators $V$ and $W$. The time at which the second term
becomes relevant gives the scrambling time
\be\label{scarmblingN}
t_*\sim\beta \log N^2\,.
\ee
The Lyapunov exponent $\lambda_L$ has a universal bound \cite{Maldacena:2015waa}
\be\label{Lbound}
\lambda_L\leq\frac{2\pi}{\beta}
\ee
and is saturated by black holes in Einstein gravity \footnote{Theories with large central charge $c$ and a sparse light spectrum also saturate this bound \cite{LargeCBound}.}. This gives support to the claim that black holes are the fastest scramblers in nature \cite{FastScrambling}. Consequently, the above bound has been used as a criterion to discriminate between CFTs that may have Einstein gravity duals \cite{ChaoticCFTs}.

An interesting question we may ask is if we can come up with other examples of systems that are maximally chaotic, i.e. that saturate the bound (\ref{Lbound}), but with no explicit gravitational degrees of freedom. In this Letter we will answer this question positively. In particular, the system we will consider is a Brownian particle (quark) coupled to a (strongly interacting) thermal plasma.

\noindent \textbf{2. Setup.}
In the context of AdS/CFT, a heavy quark in a thermal bath is dual to an open string living in a black brane geometry \cite{BrownianPapers}:
\begin{align}\label{bbmetric}
&ds^2=-r^2f(r)dt^2+{dr^2\over r^2f(r)}+r^2dx^2,\\
&f(r)=1-\left({r_H\over r}\right)^{d-1}.
\end{align}
In these coordinates, the boundary is located at $r\to\infty$. The temperature of the dual CFT corresponds to the Hawking temperature of the black brane:
\be
T=\frac{1}{\beta}=\frac{(d-1) r_H}{4\pi}\,.
\ee
In the following, we will focus on $d=3$, but the generalization to higher dimensions is straightforward.
The dynamics of an open string in such a background follows from the Nambu-Goto (NG) action:
\begin{equation}\label{NGAction}
S_{\text{NG}} = -\frac{1}{2\pi\alpha'} \int d\sigma d\tau \sqrt{-\text{det}\,\gamma_{\alpha\beta}},
\end{equation}
where $\gamma_{\alpha\beta}=g_{\mu\nu}\partial_\alpha X^\mu\partial_\beta X^\nu$ is the induced metric on the world sheet and $X^{\mu}{(\tau,\sigma)}$ are the embedding functions into the target space. We consider only the term corresponding to the tension
of the string and ignore terms which might arise from couplings to other bulk fields \footnote{We expect the coupling to other situation-specific bulk
fields to not significantly alter the results of our analysis.}.

Consider the static gauge $(\tau,\sigma)=(t,r)$ and parametrize the embedding as $X^{\mu}=\{t,r,X(t,r)\}$. The position of the quark is given by $x(t)=X(t,r_c)$, where $r_c$ is a UV cutoff.
We assume that the quark is static (in average), $\langle x(t)\rangle=0$, and consider small fluctuations due to its interactions with the thermal plasma. In the gravity side, this corresponds to studying perturbations of a static string that hangs from the boundary to the horizon, with embedding $X(t,r)=0$. Indeed, one can easily check that $X(t,r)=0$ is a solution of the NG equations of motion \footnote{In $d=3$, the metric (\ref{bbmetric}) is diffeomorphic to AdS$_3$. The coordinate transformation that maps the two brings the static solution to a solution of constant proper acceleration \cite{Acceleration}, with its two end points reaching the boundary. The induced metric in this case has a wormhole \cite{EREPR} and has been linked to the ER$=$EPR proposal \cite{Maldacena:2013xja}.}. For this solution, the induced metric on the world sheet is an AdS$_2$ black hole \footnote{The coordinate transformation $r=r_H\coth(r_H x)$ brings (\ref{inducedg}) into the more standard metric of the Schwarzschild-AdS$_2$ black hole; cf. Eq. $(3.3)$ in \cite{Spradlin:1999bn}. It will be interesting to find a connection with the SYK model \cite{SYK,Maldacena:2016hyu}.}:
\be\label{inducedg}
ds_{{\scriptscriptstyle ws}}^2=\gamma_{\alpha\beta}d\sigma^\alpha d\sigma^\beta=
                         -r^2f(r)dt^2+\frac{dr^2}{r^2f(r)}.
\ee
Thus, perturbations over this static string embedding correspond to perturbations on top of this black hole. This is the first indication that suggests the possible appearance of chaos, since black holes are known to $i)$ be fast scramblers \cite{FastScrambling} and $ii)$ saturate the bound on $\lambda_L$ \cite{Maldacena:2015waa}. However, in this setup, what plays the role of Newton's constant $G_N\sim 1/N^2$ is now $\alpha'\sim1/\sqrt{\lambda}$. Indeed, the number of degrees of freedom available is proportional to $\sqrt{\lambda}$, as can be seen, for example, from the computation of the entanglement entropy between the end points of the string \cite{EREPR,Jensen:2013lxa}:
\be\label{SEEstring}
S_{{\scriptscriptstyle EE}}=\frac{\sqrt{\lambda}}{3}\,.
\ee
In practice, in order to determine if the system is chaotic or not, we need to compute the following OTOC \footnote{The choice of operators $V$ and $W$ in (\ref{StringOTOC}) has been made in order to have a direct analogy with \cite{larkin}.}:
\be\label{StringOTOC}
\langle p\hspace{0.1em} x(t) p\hspace{0.1em} x(t)\rangle=\langle \dot{X}(0,r_c) X(t,r_c) \dot{X}(0,r_c) X(t,r_c)\rangle.
\ee
This can be obtained using standard techniques of quantum field theory in curved space, focusing on the world sheet theory (\ref{NGAction}) and regarding the embedding functions $X(t,r)$ as quantum fields \footnote{A computation of a similar four-point function in empty AdS appears in \cite{Giombi:2017cqn}.}.
\begin{figure}[t!]
\includegraphics[angle=0,width=0.35\textwidth]{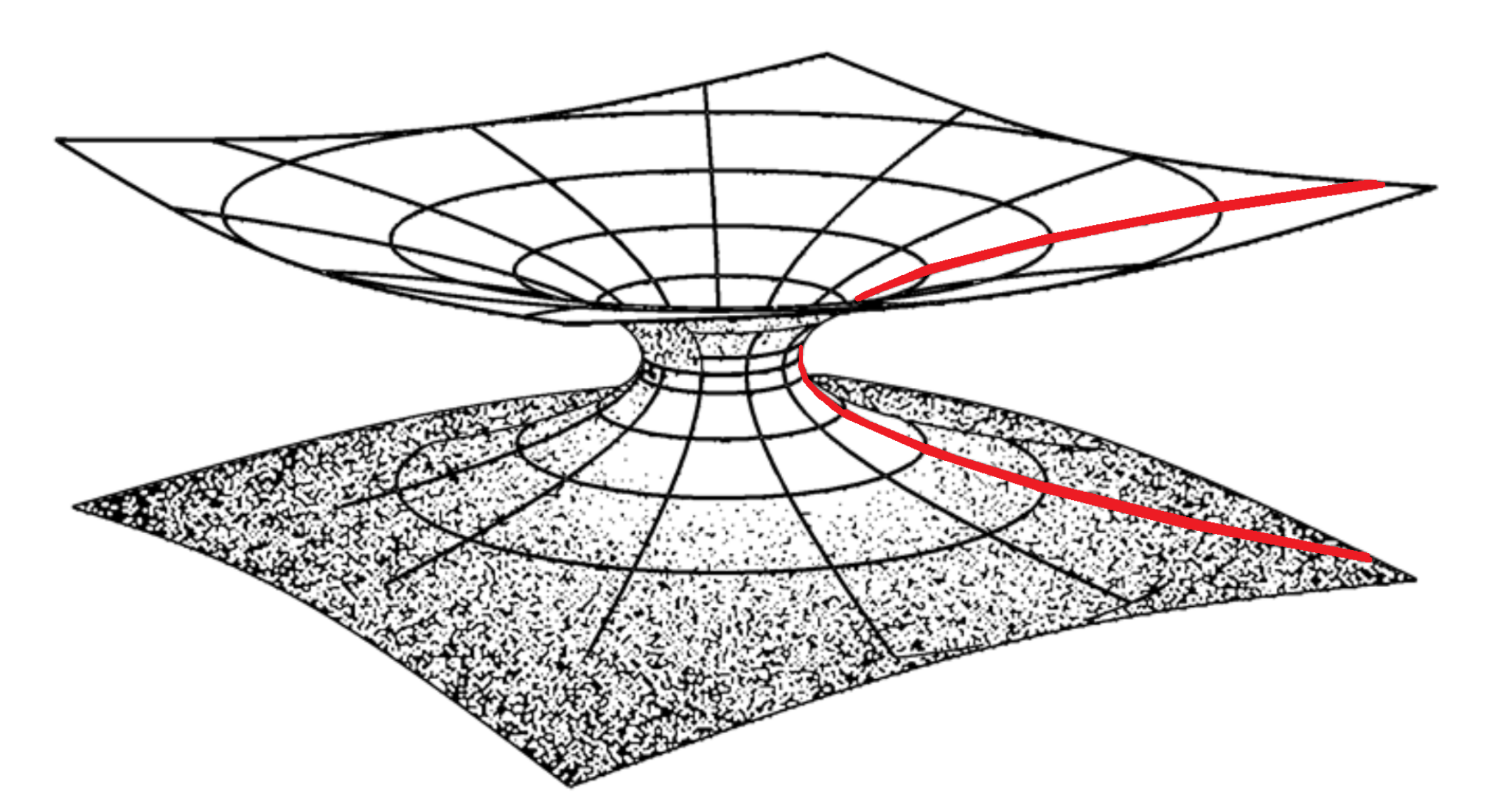}
\vspace{-2mm}
\caption{\small The setup: a string (shown in red) stretching between the two asymptotic boundaries of an eternal AdS black hole.\label{figsetup}}
\end{figure}

Before proceeding further, it will be convenient to recast the problem in terms of
Kruskal coordinates, $t=\frac{1}{2r_H}\log (-\frac{u}{v})$ and $r=r_H(\frac{2}{1+uv}-1)$, and work in the gauge $(\tau,\sigma)=(u,v)$. The string in this case stretches between the two asymptotic boundaries of an eternal AdS black hole (see Fig. \ref{figsetup}). For the static solution $X(u,v)=0$, we find that the induced metric is given by
\be
  ds_{{\scriptscriptstyle ws}}^2=-{4 du dv \over (1+uv)^2},
\ee
i.e., an AdS$_2$ wormhole. The equations for the fluctuations over this embedding follow from the action
\be\label{NGkruskal}
S_{\text{NG}} = -\frac{1}{\pi\alpha'} \int du dv \sqrt{\frac{1-r_H^2(1-uv)^2\partial_uX\partial_vX}{(1+uv)^4}}.
\ee

\noindent \textbf{3. Four-point OTOC.} In order to compute the relevant OTOC, we will use the techniques and approximations developed in \cite{Shenker:2014cwa},
adapted to the world sheet theory.

\noindent \emph{3.1 Overlapping states.}
We represent $D(\{ t_i \}) = \langle W(t_1)V(t_2)W(t_3)V(t_4) \rangle$ as the overlap of two states:
\be
  |\psi \rangle = W(t_2)^\dag V(t_1)^\dag |\Psi\rangle, \quad
  |\psi' \rangle = V(t_3) W(t_4) |\Psi\rangle,\label{states}
\ee
where $|\Psi\rangle$ is the two-sided purification: the thermofield double state. The $V$ and $W$ operators create two perturbations on the string. If the difference in times $t_2 - t_1$ and $t_4 - t_3$ are large, then the relative boosts between the wave packets are also large.

\begin{figure}[t!]
\includegraphics[angle=0,width=0.45\textwidth]{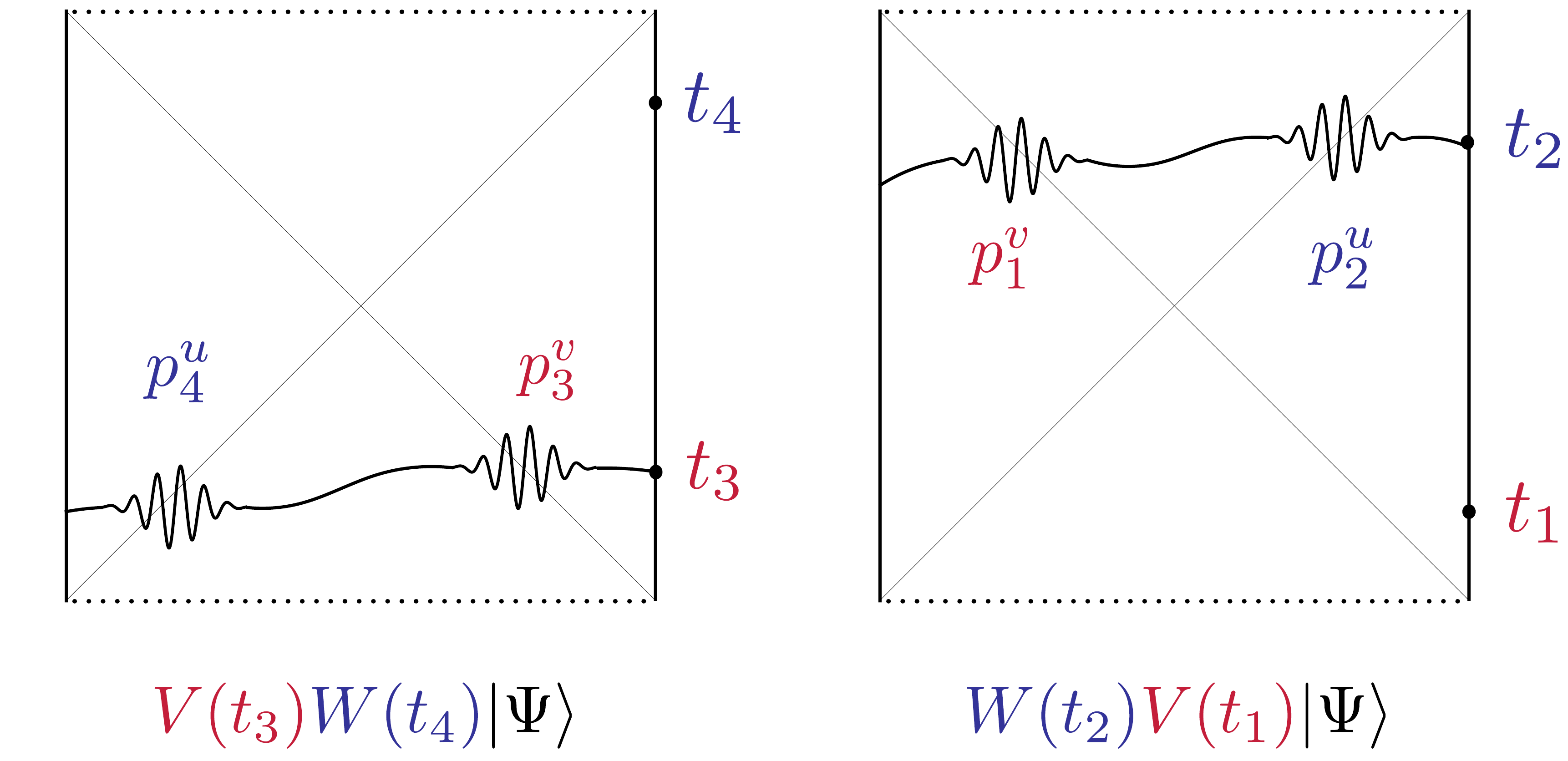}
\vspace{-2mm}
\caption{\small The four-point function (\ref{DOTOC}) as an inner product of the two states (\ref{instate}) and (\ref{outstate}). The solid lines represent spacelike slices
of the string world sheet, while the wiggles correspond to operator insertions near the horizons.
\label{fig1}}
\end{figure}

In Kruskal coordinates, the perturbation created by $W$ will have large $p^v$ and will be moving near the $u=0$ horizon. Similarly, the perturbation created by $V$ will have large $p^u$ and will be moving near the $v=0$ horizon. We represent the $W$ quantum in the Hilbert space on the
$v = 0$ horizon, and the $V$ quantum on the $u = 0$ horizon. Then $|\psi' \rangle$ is the ``in'' state
\be
  V(t_3) W(t_4) |\Psi\rangle = \int \psi_3(p_3^u) \psi_4(p_4^v) |p_3^u \, p_4^v \rangle_\textrm{in}.\label{instate}
\ee
Similarly,  $|\psi \rangle$ is an ``out'' state given by
\be
  W(t_2) V(t_1) |\Psi\rangle = \int \psi_1(p_1^u) \psi_2(p_2^v) |p_1^u \, p_2^v \rangle_\textrm{out}.\label{outstate}
\ee
The normalization of the states is given by
\be
  \langle p^v | q^v \rangle = {4 p^v \over \pi } \delta(p^v - q^v).
\ee
The wave functions can be expressed using the Fourier-transformed bulk-to-boundary propagators
\begin{align}
 & \psi_1(p^u) = \int dv e^{2i p^u v} \langle \varphi_V(u,v) V(t_1)^\dag \rangle|_{u=0},\label{Fourier1}\\
 & \psi_2(p^v) = \int du e^{2i p^v u} \langle \varphi_W(u,v) W(t_2)^\dag \rangle|_{v=0},\\
 & \psi_3(p^u) = \int dv e^{2i p^u v} \langle \varphi_V(u,v) V(t_3) \rangle|_{u=0}, \\
 & \psi_4(p^v) = \int du e^{2i p^v u} \langle \varphi_W(u,v) W(t_4) \rangle|_{v=0},\label{Fourier4}\vspace{-1cm}
\end{align}
where $\varphi_{V,W}$ are the world sheet fields dual to the operators $V$ and $W$. Finally, the four-point function is given by the overlap (see Fig. \ref{fig1} for a pictorial representation)
\be
  D = \int  d\{p_i\} \,  \psi_3^*(p_3^u) \, \psi_4^*(p_4^v) \,  \psi_1(p_1^u) \,  \psi_2(p_2^v) \, {}_{\textrm{in}}\langle p_3^u \, p_4^v | p_1^u \, p_2^v \rangle_\textrm{out}\label{DOTOC}
\ee
We still need to compute the bracket in the integrand. In the center-of-mass frame, if the relative boost is large, then the momenta $p_1^u, p_2^v, p_3^u, p_4^v$ are large and momentum conservation implies $p_1^u \approx p_3^u$, $p_2^v \approx p_4^v$. Within the two-particle Hilbert space, we can approximate
$ |p_1^u \, p_2^v \rangle_{\textrm{out}} \approx e^{i \delta(s)} |p_1^u \, p_2^v \rangle_{\textrm{in}}$,
where $s = 4 p_1^u p_2^v$ is a Mandelstam variable.

\noindent \emph{3.2 Phase shift.}
Let us now compute $\delta(s)$.
We define $R=v+u$, $T=v-u$ and rescale $X\to Xl_s/r_H$, where $l_s\equiv\sqrt{2\pi\alpha'}$.
At quadratic order, the action (\ref{NGkruskal}) reads
\be
  S_0 = {1\over 2} \int dT dR\left(  \dot X^2 - X'^2 \right).
\ee
We are interested in high-energy collisions near the horizons, so we considered the flat space approximation and set $uv=0$.
In the center-of-mass frame, the solution for two equal perturbations moving in opposite directions is
\be
  \label{eq:xwavepackets}
  {X(T,R) }  =F(T+R) + F(T-R),
\ee
where $F(\xi)$ is assumed to vanish outside a window around $\xi=0$. In this approximation, the two wave packets simply pass through each other.
 Let us now consider the subleading interacting term in the action:
\be
  S_1 = {l_s^2\over 8} \int dT dR\left(  \dot X^2 - X'^2 \right)^2.
\ee
Evaluating $S_1$ on the background of two wave packets will yield the phase shift.
By plugging in (\ref{eq:xwavepackets}), we get
\be
  \label{eq:s1}
  S_1 = l_s^2 \left( \int d\xi F'(\xi)^2 \right)^2.
\ee
Since we have $\delta(s) = S_1$, all that is left to do is to express the action in terms of the Mandelstam variable of the collision.
The target space current is given by
\be
  P^a_\mu = -\sqrt{-\gamma} \gamma^{ab} \eta_{\mu\nu} \partial_b X^\nu.
\ee
The string energy is then an integral over a spacelike slice on the world sheet
\be
  E[X] = -\int d\sigma P^\tau_T = \int d\sigma  \sqrt{-\gamma} \gamma^{\tau b} \eta_{T\nu} \p_b X^\nu.
\ee
%Here $\tau=T$ is the worldsheet time coordinate.
This gives a divergent energy for the infinitely long string
\be
  E[F] \simeq   \int dR  \left[ 1+  F'(T-R)^2+F'(T+R)^2 + \cdots\right].
\ee
The expansion in terms of $|F'| \ll 1$ is necessary, because we have neglected terms beyond $S_1$ in the expression for the phase shift.
The energy of the wave packets is thus
\be
  \Delta E = E[F] - E[0] = 2 \int d\xi F'(\xi)^2,
\ee
and the Mandelstam variable is $s = (\Delta E)^2$. Comparing this formula with (\ref{eq:s1}), we get
\be\label{phaseresult}
 \delta(s) = {s l_s^2 \over 4}.
\ee
The phase shift can also be computed from light-cone quantization for a bosonic critical string as in \cite{Dubovsky:2012wk}. It is easy to check that their method also reproduces our phase shift formula (\ref{phaseresult}).

\noindent \emph{3.3 Integral over momenta.}
The bulk-to-boundary propagator for an operator with conformal dimension $\Delta$ is
\be\label{kruskal}
\langle \phi(u,v){ \cal O}(t') \rangle =c_{ \cal O}\left(\frac{1+u v}{ u e^{t'}- v e^{-t'}+(1-uv)}\right)^{\Delta}\,,
\ee
where we have used Kruskal coordinates in the bulk and set $r_H=2\pi/\beta=1$ for simplicity. The temperature dependence can be restored by dimensional analysis whenever necessary. We evaluate these propagators at one of the horizons ($u,v=0$) and perform the Fourier transforms in \eqref{Fourier1}-\eqref{Fourier4}:
\begin{align}\label{psi1}
\psi_1(p^u,t_1)&=-2 \pi p^u \theta(p^u) e^{2\, t_1^*+2 ip^u e^{t_1^*}},\\
\psi_2(p^v,t_2)&= i\pi p^v \theta(p^v) e^{-\, t_2^*-2 ip^v e^{-t_2^*}},\\
\psi_3(p^u,t_3)&=-2 \pi p^u \theta(p^u) e^{2\, t_3+2 ip^u e^{t_3}},\\
%\end{align}
%\begin{align}
\psi_4(p^v,t_4)&= i\pi p^v \theta(p^v) e^{-\, t_4-2 ip^v e^{-t_4}}\label{psi4},
\end{align}
where the complex conjugate in $t_1$ and $t_2$ appears because we are considering Hermitian conjugates for the first and the second propagators in $|\psi\rangle$. Since $W(t)=x(t)$ and $V(t)=\dot x(t)$, we have already set $\Delta_W=1$ and $\Delta_V=2$. With \eqref{psi1}-\eqref{psi4} and $\delta=p^u p^v \,l_s^2$,  we are ready to perform the overlap integration (\ref{DOTOC}).
By changing the variables $p^u=-\frac{p}{2 i \left(-e^{t_1}+e^{t_3}\right)}$ and $p^v=\frac{q}{2 i \left(-e^{t_2}+e^{t_4}\right)}$ and fixing the end points as $t_1=i\epsilon_1,\,t_2=t+i\epsilon_2,\,t_3=i\epsilon_3,\,t_4=t+i\epsilon_4$, we find
\begin{align}
\nonumber
&\langle V(i\epsilon_1)W(t+i\epsilon_2)V(i\epsilon_3)W(t+i\epsilon_4)\rangle =\\
&\qquad\qquad\qquad\qquad C \int_{0}^{\infty} dp dq \,p^3 q \,e^{-p-q} e^{i l_s^2 e^t p q/4 \epsilon_{13}\epsilon_{24}^{*} }, \label{4point}\\
\nonumber
&\langle V(i\epsilon_1)V(i\epsilon_3)\rangle\,\langle W(t+i\epsilon_2)W(t+i\epsilon_4)\rangle =\\
\label{4pointdis}
&\qquad\qquad\qquad\qquad C \int_{0}^{\infty} dp dq\, p^3 q \, e^{-p-q},
\end{align}
where we have defined the constants
\begin{align}
&C\equiv c_V^2c_W^2\frac{\pi^4}{1024}\csc ^4\left(\frac{\epsilon_1-\epsilon_3}{2}\right) \csc ^2\left(\frac{\epsilon_2-\epsilon_4}{2}\right),\\
&\epsilon_{ij}\equiv i(e^{i\epsilon_i}-e^{i\epsilon_j}).
\end{align}
The integrals above can be computed exactly in terms of the exponential integral, $Ei(z)=-\int^{\infty}_{-z}e^{-t} dt$. However, the results can be trusted only up to $\mathcal{O}(l_s^2)$, since we truncated the action at this order. Performing this approximation, the normalized four-point function reads
\be
\frac{\langle V(i\epsilon_1)W(t+i\epsilon_2)V(i\epsilon_3)W(t+i\epsilon_4)\rangle}{\langle V(i\epsilon_1)V(i\epsilon_3)\rangle\,\langle W(t+i\epsilon_2)W(t+i\epsilon_4)\rangle}
\simeq1+\frac{2i l_s^2 e^{t}}{\epsilon_{13}\epsilon_{24}^{*}}.
\ee

\noindent \emph{3.4 Scrambling and Lyapunov exponent.}
Although we denoted the imaginary time parameters as $\epsilon_i$, they do not necessarily have to be small. For instance, if we subtract
$\beta/2$ from $\epsilon_1$ and add the same to $\epsilon_4$, we can obtain two-sided correlators from the above one-sided expectation value.
A canonical choice made in \cite{Maldacena:2015waa} consists in setting $\epsilon_1=\beta/2=\pi$, $\epsilon_2=-\beta/4=-\pi/2$, $\epsilon_3=0$, $\epsilon_4=\beta/4=\pi/2$. This corresponds to the insertion of the $V$ and $W$ operators at equal spacing around the thermal circle. With this choice, one gets $\epsilon_{13}=-2i$ and $\epsilon_{24}=2$. Finally, by restoring the temperature dependence, we find that the four-point function (\ref{otoc}) is given by
\be
f(t)=1-\frac{\pi}{\sqrt{\lambda}}e^{2\pi t/\beta}.\label{fstring}
\ee
The above equation must be contrasted with the result for correlator in the pure gravity sector (\ref{fgravity}).
From (\ref{fstring}), we can read off the Lyapunov exponent
\be
\lambda_L=\frac{2\pi}{\beta},
\ee
which saturates the bound (\ref{Lbound}), and the scrambling time
\be\label{scarmblinga}
t_*\sim\beta \log\sqrt{\lambda}.
\ee
Thus, even though the world sheet theory is not gravitational, it is maximally chaotic and exhibits the fast scrambling property of black holes. In addition, there is also a parametrically large hierarchy between scrambling and dissipation determined, in this case, by the small parameter $\alpha'\sim1/\sqrt{\lambda}$ instead of the standard $G_N\sim1/N^2$. The fact that $t_*$ scales the way it does can be easily understood, since $\sqrt{\lambda}$ is proportional to the excess of entropy due to the probe string (\ref{SEEstring}), and these are precisely the degrees of freedom that are being scrambled.

\noindent \textbf{4. Complexity.} Black holes are known to excel at another information theoretic task,
namely, the processing of information. The rate of quantum information processing is measured by computational complexity $\mathcal{C}$. Complexity
counts the minimal number of gates needed to build a quantum circuit which
prepares the state from a particular reference state. It grows linearly at late times and obeys the bound $d\mathcal{C}/dt\leq2E/\pi\hbar$ \cite{Lloyd}.  It is then
interesting to ask about complexity in our present setup.

In AdS/CFT, there are two proposals to compute complexity, the Complexity$=$Volume \cite{CV} and the Complexity$=$Action \cite{CA} conjectures, both satisfying the bound.
In the former, the complexity is proportional to the spatial volume of the Einstein-Rosen bridge $V$:
\be
\mathcal{C}\sim \frac{V}{G_N\ell},
\ee
where $\ell$ is some length scale. This quantity does indeed grow linearly over time at late times. In order to compute the correction to
$\mathcal{C}$ due to the probe string, one would need to consider the backreaction of the string on the bulk geometry
and compute the new volume $V$. Here we proceed differently. We define the dimensionless quantity
\be
\mathcal{C}_{{\scriptscriptstyle ws}}= \frac{\tilde{\ell}\hspace{0.1em} V_{{\scriptscriptstyle ws}}}{\alpha'},
\ee
as a ``world sheet complexity'', where $\tilde{\ell}$ is a time scale and $V_{{\scriptscriptstyle ws}}$ is the length of the world sheet wormhole. A brief computation yields at late times
\be
\frac{dV_{{\scriptscriptstyle ws}}}{dt}=\frac{2\pi}{\beta}\quad\rightarrow\quad \frac{d\mathcal{C}_{{\scriptscriptstyle ws}}}{dt}=\text{const}.
\ee
It would be interesting to ask about its significance in the CFT language.
The second proposal for complexity gets rid of the arbitrary length scale $\ell$ and states that
\be\label{WdWcomplex}
\mathcal{C}\sim \frac{\mathcal{S}_\text{WDW}}{\pi\hbar},
\ee
where $\mathcal{S}_\text{WDW}$ is the bulk action evaluated on the Wheeler-DeWitt patch. The correction of
$\mathcal{C}$ due to the probe string in this case is simpler: it is given by the NG action evaluated on the
Wheeler-DeWitt patch \cite{flavorcomplexity}. Notice that if we were to define a world sheet complexity using the
world sheet geometry, the result would be equivalent to (\ref{WdWcomplex}). However, at least in $d=3$ we find that there is an ambiguity on defining the Wheeler-DeWitt patch, because the maximally extended world sheet geometry gets past the $uv=1$ edges (cf. Fig. 5 in \cite{Hubeny:2014kma}). We hope to come back to this point in the future.

\noindent \textbf{5. Discussion.} We have presented the first example of a nongravitational system
that is maximally chaotic, i.e., that saturates the universal bound on the Lyapunov exponent (\ref{Lbound}).
The other two known examples that saturate the bound AdS black holes in Einstein gravity \cite{Maldacena:2015waa} and the SYK model \cite{SYK}, which contains an AdS$_2$ dilaton gravity sector \cite{Maldacena:2016hyu}. Even though the world sheet theory does not contain gravitational degrees of freedom, it is worth recalling that the world sheet theory of strings shares some interesting similarities with
theories of quantum gravity, including the absence of local off-shell observables, a minimal length, a maximum achievable (Hagedorn) temperature, as well as (integrable relatives of) black holes \cite{Dubovsky:2012wk}.

In summary, the maximal chaotic exponent for the string follows from the following two points: $i)$ the induced world sheet metric has a horizon;
therefore, by Rindler kinematics, the relation between world sheet scattering energy and time is $s\sim e^{2\pi t/\beta}$. And $ii)$ the eikonal phase
is $\delta\sim\alpha' s^p$ with $p=1$. This result is quite nontrivial. In ordinary QFT, a spin $J$ field exchanged in the Mandelstam $t$ channel gives $p=J-1$.
Causality and unitarity further constrain the value of the exponent to be $p\leq1$, since $e^{i\delta(s)}$ must be analytic in the upper half of the complex
$s$ plane and $|e^{i\delta(s)}|\leq1$ \cite{Camanho:2014apa}. The NG theory has infinitely many higher derivative nonrenormalizable terms that appear nonlinearly in the action. As  explained  in  \cite{Zamolodchikov:1991vx},  the  requirements  of  unitarity,  crossing  symmetry,  and  analyticity
restrict the phase shift to take the form
\be
e^{i2\delta{(s)}}=\prod_j\frac{\mu_j+s}{\mu_j-s}e^{iP(s)}\,,
\ee
where $P(s)$ is an odd polynomial in $s$ and $\mu_j$ are located in the lower half of the complex
plane and either lie on the imaginary axis or come in pairs symmetric with respect to it. What is surprising is that
the $\mu_j$ and $P(s)$ for the NG theory conspire to give the required phase shift mimicking a single graviton
exchange (see also \cite{Dubovsky:2012wk}).

Finally, let us comment on the extension of the chaos bound conjectured in \cite{Maldacena:2015waa} to our setup. The proof of the maximal Lyapunov exponent
relies on two points. The first one is a result bounding the derivative of any function, which was shown to hold in general. The second one
is the assumption that the error $\varepsilon$ \footnote{Properly defined in \cite{Maldacena:2015waa}.} of the late-time factorization of the OTOC is small. A quick calculation shows that in our setup $\varepsilon\sim1/\sqrt{\lambda}$, which holds true as long as $\lambda$ is large. This, together with the fact that $\delta(s)\sim s$ seems to hold beyond leading order \cite{Dubovsky:2012wk}, suggests that perturbative higher-order $\alpha'$ corrections should respect the bound, although it would be interesting to see an explicit calculation. Furthermore, at weak 't Hooft coupling, the strength of scattering in the
gauge theory is of the order of $\lambda$, so one would expect $\lambda_L\sim\lambda/\beta$, parametrically smaller than the strong coupling result.

There are a few directions that may be worth exploring in the future. Two interesting generalizations to consider are $i)$ a higher-dimensional target space and $ii)$ higher-dimensional probes in the bulk such as $Dp$-branes. From the latter,
one could also compute the associated butterfly velocities and compare with the charge diffusion results \cite{Diffusion}. One could also
repeat the computations of the OTOC presented here in more complex shockwave geometries, i.e., segmented strings in AdS space \cite{Segmented}.
Finally, it would be interesting to understand whether the chaotic behavior observed here can emerge in
other field theories in a black hole background or whether there are specific features of the string action that make it chaotic.

%%%%%%%%%%%%

\noindent \emph{Note added:} Recently, we became aware of \cite{Murata:2017rbp} whose results overlap with ours.

\noindent \emph{Acknowledgements.}
We are grateful to M. Chernicoff, V. Hubeny, and D. Stanford for discussions and useful correspondence. This work is supported by the Netherlands Organisation for Scientific Research (NWO) and the Delta-Institute for Theoretical Physics ($\Delta$-ITP).

%\appendix
%
%\section{Appendix}

\end{document}